\documentclass[twoside,11pt]{article}

\usepackage{obs_study_style}
\usepackage{graphicx}
\usepackage{amsmath}
\usepackage{subcaption}
\usepackage{color} 
\usepackage{tikz}

\ShortHeadings{Distinguishing amongst classes of dynamic causal effects}{Steele, et al.}
\firstpageno{1}
\heading{}{}{}{}{}{Russell Steele, Naftali Weinberger, Tess Baker, and Ian Shrier}
\firstpageno{1}

\title{Modifying causal models to distinguish between transient and lasting causal effects}

\author{\name Russell Steele \email russell.steele@mcgill.ca \\
       \addr Department of Mathematics and Statistics\\
       McGill University\\
       Montr\'{e}al, Qu\'{e}bec, Canada 
       \AND
       \name Naftali Weinberger \email naftali.weinberger@gmail.com\\ 
       \addr Munich Center for Mathematical Philosophy\\
             Ludwig-Maximilians-Universität, München\\
             Munich, Germany
       \AND
       Tess Baker \email
       tess.baker@mail.mcgill.ca \\
       \addr Department of Mathematics and Statistics\\
       McGill University\\
       Montr\'{e}al, Qu\'{e}bec, Canada
       \AND
       \name Ian Shrier \email ian.shrier@mcgill.ca \\
       \addr Centre for Clinical Epidemiology \\
       Lady Davis Institute for Medical Research, Jewish General Hospital\\
        Montr\'{e}al, Qu\'{e}bec, Canada
       }

\begin{document}
\maketitle

\begin{abstract}
This paper considers how to classify the effects of interventions in causal models for outcomes and exposures observed over time. First, we demonstrate the limitations of the most common uses of potential outcomes and causal directed acyclic graphs (DAGs) for capturing all possible interventions in a time-varying framework, particularly in problems where the key question concerns interventions to maintain or change equilibrium behaviour. Second, we adopt a system-and state-based approach rather than a measurement-based approach to identify the causal parameters. In particular, we discuss how assumptions about the system's equilibrium and the effects of interventions on that equilibrium can allow for more specific causal interpretations and clarify the goals of design and analysis.  Third, we show how the ability to identify the  the causal parameters of a time-varying system depends on the selection of timepoints for measuring the system's states. We address this by proposing a novel version of the null-effect, which is designed to distinguish between transient and lasting causal effects.


\end{abstract}

\section{Introduction \label{sec:intro}}

The development of potential outcome models and the use of causal directed acyclic graphs (DAGs) in causal inference for observational data has changed the landscape of modern casual inference over the last 30 years. Researchers have used these models and associated estimation methods with great success in the analysis of cross-sectional and longitudinal data in public health, economics, education and many other areas where the effects of interventions cannot always be evaluated in designed experiments. Despite the successes, there are important causal questions that cannot be easily addressed by current causal inference methods and approaches, in particular when they are applied to systems with complex temporal dynamics.   

Consider estimating the efficacy of a new chemotherapeutic intervention to reduce breast cancer tumour size improve survival in a randomized controlled trial. Using methods consistent with common practice for causal DAGs in health research, our causal DAG would include a node for treatment, a node for tumour size, a node for survival, and possibly nodes for strong prognostic factors. Our randomized trial study would select patients from a well-defined cancer population and include a clear standard treatment. We would record and compare the proportion of positive discrete outcomes (remission, survival) or summaries of continuous outcomes (tumour size reduction, quality of life changes). We would consider the treatment efficacious or not depending on the magnitude and uncertainty of the estimated differences. If we were interested in effects at different time points, we would include time-dependent nodes in our causal DAG (e.g. dynamic treatment regimes), or nodes that represent changes in distributions between timepoints instead of absolute distributions at the timepoints \citep{glymour2005baseline}. In an observational setting, estimation methods would change due to the possible presence of confounding, but all other principles would remain the same. 

Although these methods provide important insights into treatment efficacy over time, all inferences are restricted to the timing of the measurements actually chosen, because there may be different ``causal effects'' depending on the duration of the intervention, the lag time between intervention initiation and treatment effect, and the duration of the treatment effect. These issues are especially relevant in cases where an intervention leads to a transient change in the outcome followed by the outcome returning to its baseline value (or ``equilibrium'' in the general sense of the word). This points to a limitation of standard definitions of the null effect, which fail to differentiate between transient outcome effects where the system returns to equilibrium, and permanent outcome effects that do not subside. 

Whereas the variables in the mainstream causal models refer to measurements at particular time points (i.e. values of the variables), it is also possible to incorporate the dynamics of the system. By ``dynamics'', we mean that some interventions could also affect the magnitude of the causal relationships between the measured variables (i.e. the parameters in the causal equations linking variables at different timepoints). Differential equations are common in science for representing dynamical systems  mathematically, and causal diagrams that include such equations are known as ``dynamic causal models'' (as contrasted with ``mainstream causal models"). These differential equations lead to trajectories that help us distinguish whether an intervention will have a transient or long-term effect. 

The purpose of this paper is to show how comparing common and dynamic causal models for the same context / system leads to new insights. First,we show that causal effects on the system dynamics cannot be achieved by interventions that simply change the values of variables (which we refer to as ``states'') on the causal DAG, but require interventions on the dynamics. Second, we offer a novel version of the null effect that distinguishes between null transient and null long-term effects which allows one to differentiate formally between the two types of interventions. In the current paper, we will focus only on the causal identification problem for dynamic systems, rather than the estimation of the defined causal effects from observed data.  

Within mainstream causal frameworks, interventions on dynamics correspond to interventions on causal parameters, rather than on the distribution of variables. Although interventions on parameters may seem foreign to many causal practitioners, we argue that they are unproblematic, and can be straightforwardly understood within other frameworks (see Section \ref{sec:intervene} for an example using Rothman's sufficient causal set framework \citep{rothman1995causes}). Our work synthesizes ideas from mainstream causal DAGs \citep{pearl2009causality} with recent work on dynamic causal models \citep{blom2023causality,weinberger2023intervening,rubenstein2018deterministic} and much earlier results on causal ordering at equilibrium \citep{iwasaki1994causality}. This leads to a new structure that improves the specificity of interventions, study design, and data collection so that we can better answer existing questions and introduce the possibility of new ones.

The next section begins by explaining that mainstream causal models are {\em state-focused}, using either DAGs or  potential outcomes for outcome causal estimation. In the third section, we use an example of cancer cells
to illustrate the limitations of the mainstream causal approach under what are typical causal interventions, and discuss challenges with estimation in a context where the theoretical model is not well-understood. We then rely on work from the philosophy community to characterize the challenges of this common {\em state-focused} model, and to provide guidance for harmonizing mainstream causal and dynamic models. 
 In the fourth section, we discuss the implications for the design and analysis of experimental and observational studies, particularly with respect to the importance of the timing of measurements.

\section{Motivating Example for Health Research}

Dynamic causal models are currently used primarily in areas that explore equilibrium behaviour for weather, chemical reactions, and so on. However, most human diseases can also be reframed as questions related to achieving or moving away from equilibrium. For example, cancer cells are continuously occurring in the human body, but are killed by the natural killer cells of our immune system \citep{mace2023human}. We are given a diagnosis of cancer when the volume of cancer cells achieves a particular threshold because the creation of new cancer cells exceeds the destruction of cancer cells, i.e. when the system is no longer in equilibrium. In studies using mainstream causal models, the analysis would focus on remission or tumour size (i.e. tumour volume) over a specific time interval but not on the shift away from equilibrium. More insights would be gained by exploring the change in tumour volume over time, which might also provide insight into whether that benefit would likely continue after the study. While one could design a study with multiple timepoints and report how the tumour volume changed over time, the timing and frequency of the 
measurements would determine exactly what could be learned and how much uncertainty might remain. Relatedly, one would not
understand what the intervention has changed in the physiological system, only that observed quantities had been changed. 

\citet{cho2023designing} provide a concrete example, which can be used to illustrate the limitations of the classical potential outcomes approach for capturing the full range of possible causal interventions. The authors investigated the structural identifiability of a dynamic model for tumour growth when there are multiple types of cancer cells present.  In particular, they examined the ability of the Lotka-Volterra competition model to produce three different kinds of cell line interactions with respect to tumour growth: competitive (both cell lines inhibit the other's growth), mutualistic (both cell lines promote the other's growth), and antagonistic (one cell line inhibits the other's growth, and the other cell line either promotes the other's growth or has no effect).  We will use the same notation and description from \citet{cho2023designing}. That is, we will assume that we are interested in characterizing the change in tumour volume (i.e. growth) where the tumour is comprised of two cell types: (1) aggressively growing cells that we would like to target if we could ($Type_{target}$) and (2) proxy cells ($Type_{proxy}$) that are less aggressive, which we can intervene upon easily, but which have only an indirect effect on the target cells. 

The dynamic system is defined by the equations 1 and 2. For the illustrations that follow, we assume that the average population behaviour can be described with the following system of ordinary differential equations,

\begin{align}
\frac{dV_{target}}{dt} & = r_{target} V_{target}(t) \left ( 1 - \frac{V_{target}(t) + \gamma_{proxy} V_{proxy}(t)}{K_{target}} \right )  \label{eq:LVeq_int_A}; &\\ 
\frac{dV_{proxy}}{dt} & = r_{proxy} V_{proxy}(t) \left ( 1 - \frac{V_{proxy}(t) + \gamma_{target} V_{target}(t)}{K_{proxy}} \right )  \label{eq:LVeq_int}. 
\end{align}
\vspace{0.25in}

\begin{figure}
    \centering
    \includegraphics[width=1\linewidth,scale=0.3]{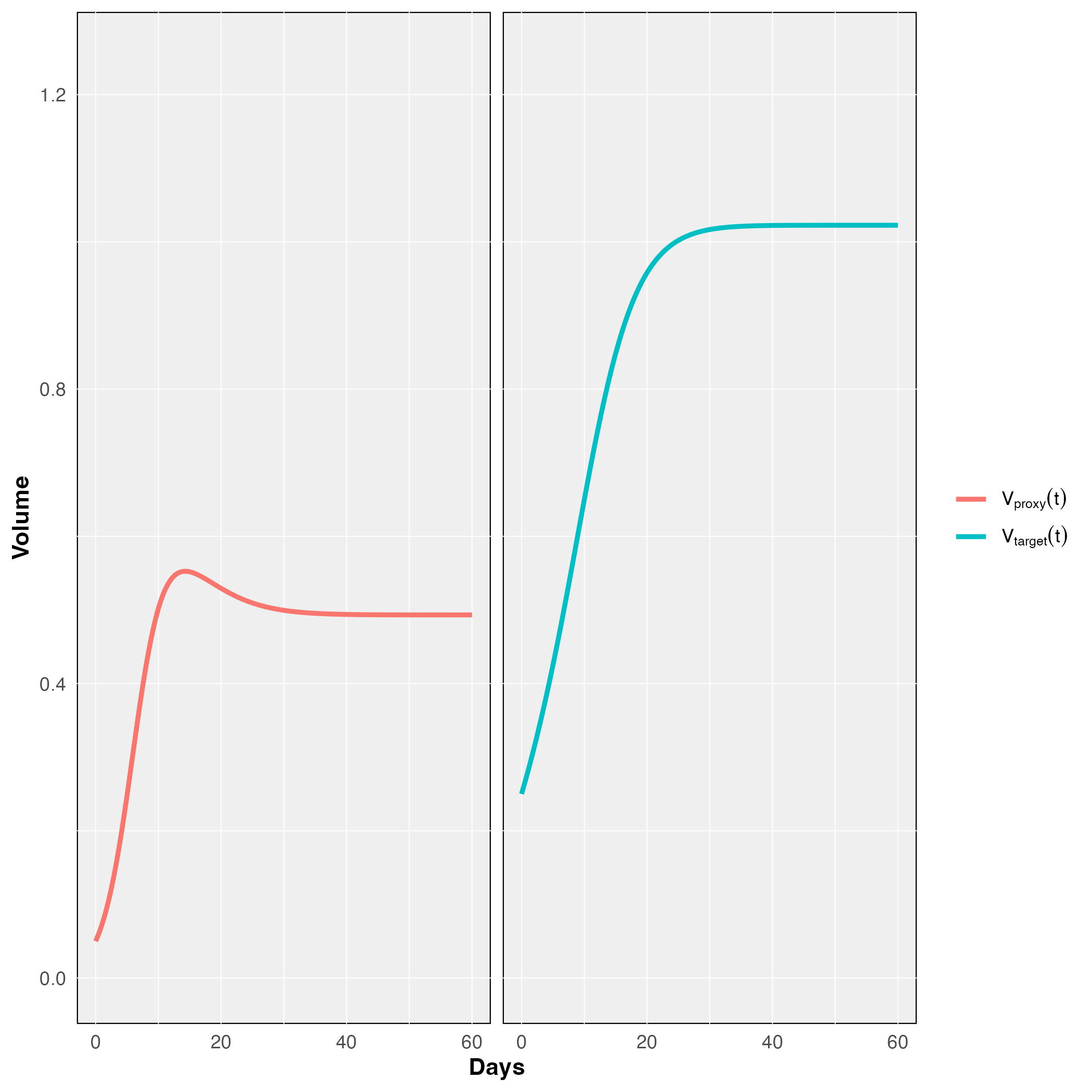}
    \caption{\label{fig:comp_noint} $Type_{proxy}$ (left) and $Type_{target}$ (right) cell volumes under the model of Equations~\ref{eq:LVeq_int_A} and \ref{eq:LVeq_int} with parameter values $r_{target}=0.16$,   $r_{proxy}=0.45$, $K_{target}=0.85$, $K_{proxy}=0.80$, $\gamma_{target}=-0.35$, and $\gamma_{proxy}=0.30$.}
\end{figure}

\noindent $V_{target}(t)$ and $V_{proxy}(t)$ are the volumes of the the two cell types as a function of $t$, $K_{target}$ and $K_{proxy}$ represent the maximum tumour volume the system can have (generally $K$ is assumed to be much larger than any particular $V_{proxy}(t)$ or $V_{target}(t)$ of interest), $r_{target}$ and $r_{proxy}$ are the exponential growth rates, and $\gamma_{proxy}$ and $\gamma_{target}$ determine the kind (i.e. competitive, mutualistic, antagonistic) and magnitude of the interaction between the target and proxy cell types. Figure~\ref{fig:comp_noint} illustrates the behaviour of the system  with initial values $V_{proxy}(0) = 0.05$, $V_{target}(0) = 0.25$, $r_{target}=0.16$,   $r_{proxy}=0.45$, $K_{target}=0.85$, $K_{proxy}=0.80$, $\gamma_{target}=-0.35$, and $\gamma_{proxy}=0.30$. 

In the next section, we characterize some issues with common potential outcome modelling for defining interventions in such a dynamic system. 

\section{Characterizing Interventions on System States \label{sec:charact}}

\subsection{Interventions on a single dynamic process}

In the most basic causal formulation for intervention (or exposure) and outcome measurements,
we can define the potential outcome for the outcome measurement as $Y^{(a)}$, 
i.e. the value that the outcome $Y$ would take on if a point exposure 
(or intervention) $A$ were taken.  Assuming only two values for $A \in \{0, 1\}$ we
could then define the causal effect of intervention $A$ in the individual as  
$Y^{(1)} - Y^{(0)}$ and then define an average causal effect in the population as:
$E(Y^{(1)} - Y^{(0)})$ under a parametric, semi-parametric or finite population 
model for the pair $\{Y^{(0)}, Y^{(1)}\}$. 

In the dynamic setting described in this paper, we need to generalize
and provide more careful indexing with respect to time. 
We first consider a single intervention $a_t$ where $a_t$ in $\{0, 1\}$ at time $t$ only affects the $V_{target}$. For the purposes of illustration, in this first example we assume the two cell types do not interact, i.e. $\gamma_{proxy}$ = $\gamma_{target}$ = $0$. Therefore, we can ignore $V_{proxy}(t)$ in this example. Initial volume is represented by $V_{target}(0)$. For example, if $a_t = 0$, then $V_{target}(t)^{a_t=0}$ is the solution of the differential 
equation for the volume of $Type_{target}$ tumour cells at time $t$ using the original $V_{target}(0)$ as the $Type_{target}$ cell starting volume and no intervention. When 
$a_t = 1$, then the value of $V_{target}(t)$ is increased (or decreased) relatively by some amount 
$\beta$, i.e. $Y^{a_t = 1}_{t} =\beta Y^{a_t=0}_{t}$. 

\begin{figure}
    \includegraphics[width=1\linewidth]{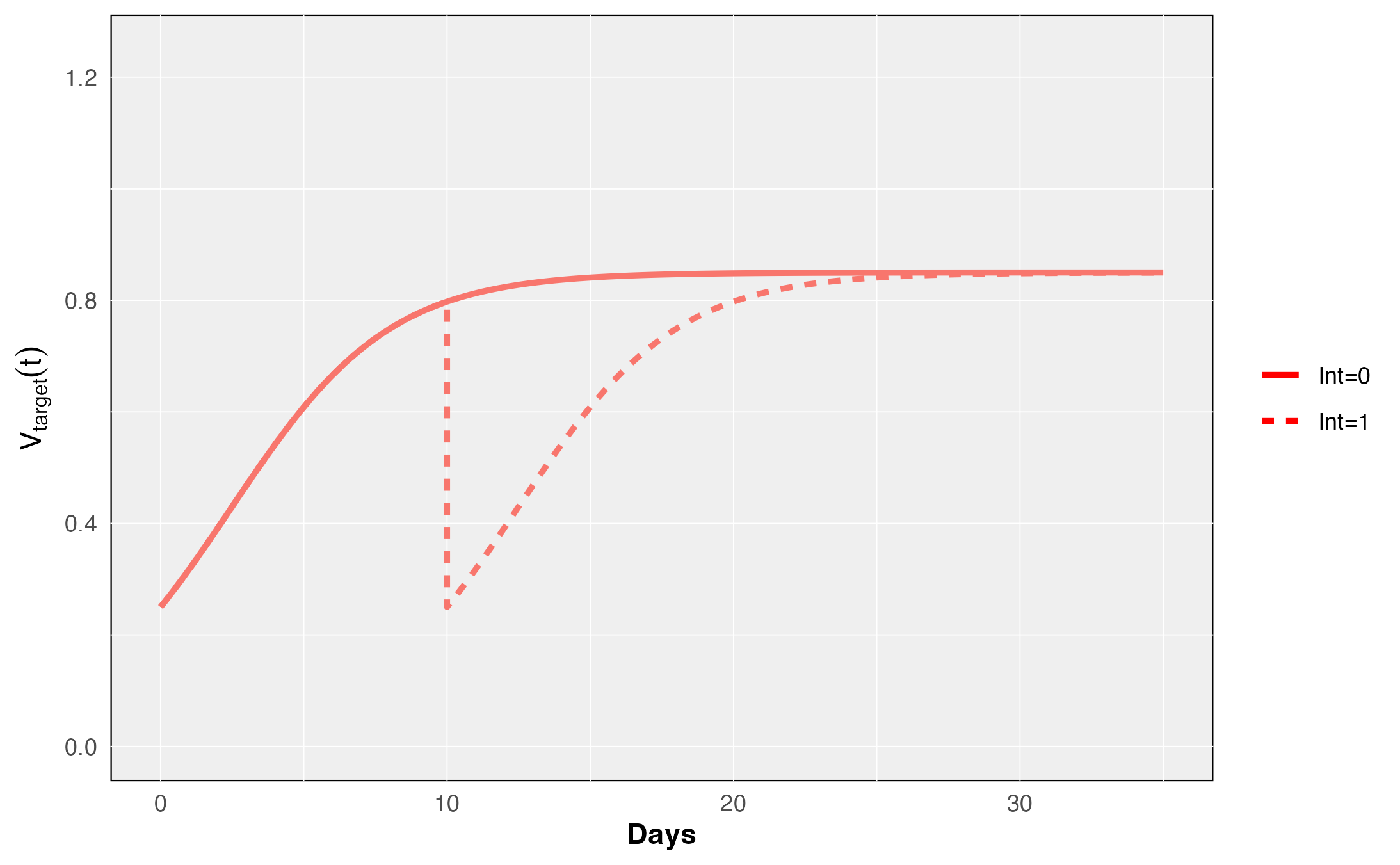}
    \caption{\label{fig:iso_int} Example of a single intervention on the $V_{target}(t)$ state at 
    time $t=10$.  The dashed line indicates the trajectory of $V_{target}(t)$ post-intervention, the solid line indicates the value of $V_{target}(t)$ with no intervention. The value of $V_{target}$ with the intervention at any time $t >10$ is equal to the value of $V_{target}$ without an intervention at time $t-10$.}
\end{figure}

Figure~\ref{fig:iso_int}
illustrates the effects of intervening directly on the target cell in the case when the target and proxy cell volumes change in isolation of each other. The effect of an intervention at time $t=10$ on $V_{target}(t)$ is to 
decrease the volume by 60\% (dashed line) relative to the value under no intervention at time 
$t=10$ (solid line). Note that $V_{target}(t)^{a_t=1} = V_{target}(t-10)^{a_t=0}$ for $t \geq 10$, which 
follows along the work of Robins \citep{robins2004optimal} and others (e.g. \citet{murphy2003optimal},  \citet{saarela2015predictive}) on state-focused causal modeling of exposures. The causal effect of the intervention in our case is not a
single value, but rather a function of time \citep[cf.][]{ecker2024causal}.  
In our example, as time increases, the effect of 
the intervention on $V_{target}(t)$ diminishes, i.e. $|V_{target}(t)^{a_t=1} - V_{target}(t)^{a_t=0}|$ 
decreases until it reaches the same equilibrium value as under no intervention, 
just at a later time point. Crucially, because a mainstream causal effect is defined at a single 
time-point, it is inexorably tied to the time-point which the measurement is taken. 

\begin{figure}
    \centering
    
\begin{tikzpicture}[
    >=stealth,
    node distance=3cm,
    dagnode/.style={circle, draw, minimum size=1cm, font=\small}
]

\node[dagnode] (V_T0) {$V_{target}(0)$};
\node[dagnode, right of=V_T0] (V_T5) {$V_{target}(5)$};
\node[dagnode, draw=red!100, thick, right of=V_T5] (V_T10) {$V_{target}(10)$};
\node[dagnode, draw=red!100, thick, above of=V_T10] (A_T10) {$A_{Int}(10)$};
\node[dagnode, draw=red!100, thick=10,dashed, right of=V_T10] (V_T15) {$V_{target}(15)$};
\node[dagnode, draw=red!100, thick=10,dashed, right of=V_T15] (V_T20) {$V_{target}(20)$};

\draw[->] (V_T0) -- (V_T5);
\draw[->,draw=red!100,] (A_T10) -- (V_T10);
\draw[->] (V_T10) -- (V_T15);
\draw[->] (V_T15) -- (V_T20);


\end{tikzpicture}

    \caption{    \label{fig:s_dag_iso} Causal Directed Acyclic Graph for $V_{target}(t)$ at different timepoints based on a single intervention on $V_{target}(t)$ at $t=20$. The solid red node at $t=10$ is intervened upon, the dashed nodes indicate the nodes after $t=10$ that are changed due to the indirect effects of the intervention. The intervention on $V_{target}(5)$ will not immediately affect $V_{target}$ at $t=10$ so the edge is removed.}

\end{figure}
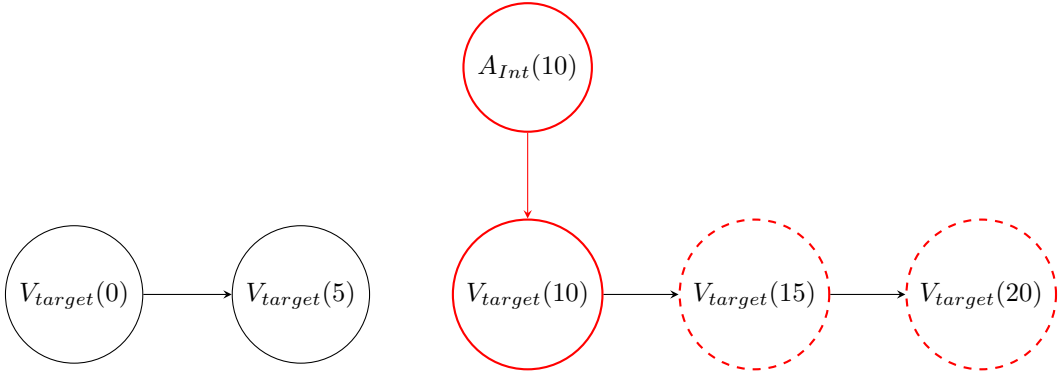

In this example, the mainstream causal DAG approach 
can capture some (but not all) of the nuance of the model dynamics. 
The mainstream causal model for this problem (Figure~\ref{fig:s_dag_iso}) contains variables for each of the 
observed time points. An intervention $A$ at 
time $t=10$ would have a direct effect on $V_{target}(10)$, and would 
have a diminishing indirect effect on $V_{target}(t)$ as time increased. One implication of this is that once one reaches a time-point at which the effect has subsided (to some arbitrary level of precision) the model will violate the causal Faithfulness condition \citep{spirtes2000causation}, which implies that an effect and its cause will be probabilistically dependent. Although our model is not itself probabilistic, one can easily see that after a certain number of time-steps the value of the target is exactly the same as it would have been in the absence of an intervention, and thus would be probabilistically independent of the intervention in a probabilistic analogue of our model incorporating random error terms at each time-step. The causal Faithfulness condition arises in constraint-based methods for causal discovery, and its failure here might just be one more reason to reject it in dynamical contexts. But the failure also  points towards a broader limitation of the DAG, which is unable to differentiate transient from permanent causal effects.

The simple example from Figure~\ref{fig:iso_int} shows an important limitation 
when formulating the effects of causal interventions in terms of changes to 
observable quantities and potential outcome states (i.e. $Y_t$ and $Y_t^{a_t}$). 
The intervention in this example does not change the equilibrium behaviour or the dynamics of 
the system; it only temporarily affects the state.

\subsection{Characterizing interventions on system dynamics }

Now consider the case where the rate of change in volume for each cell type depends on the current volume of the other cell type, as outlined in equations 1 and 2 when $\gamma_{proxy}$ and $\gamma_{target}$ are both $>0$. This type of interaction can either be competitive (both cell types inhibit the other because they are constrained by the same resources), mutualistic (each cell type increases the other because they both create a beneficial environment) or antagonistic (one cell type creates a toxic environment for the other thereby interfering with its growth, and the other cell type either has none or a beneficial effect on the first cell type). In the following example we set $\gamma_{proxy} = 0.4$ and $\gamma_{target} = 0.6$, which would indicate a competitive environment.

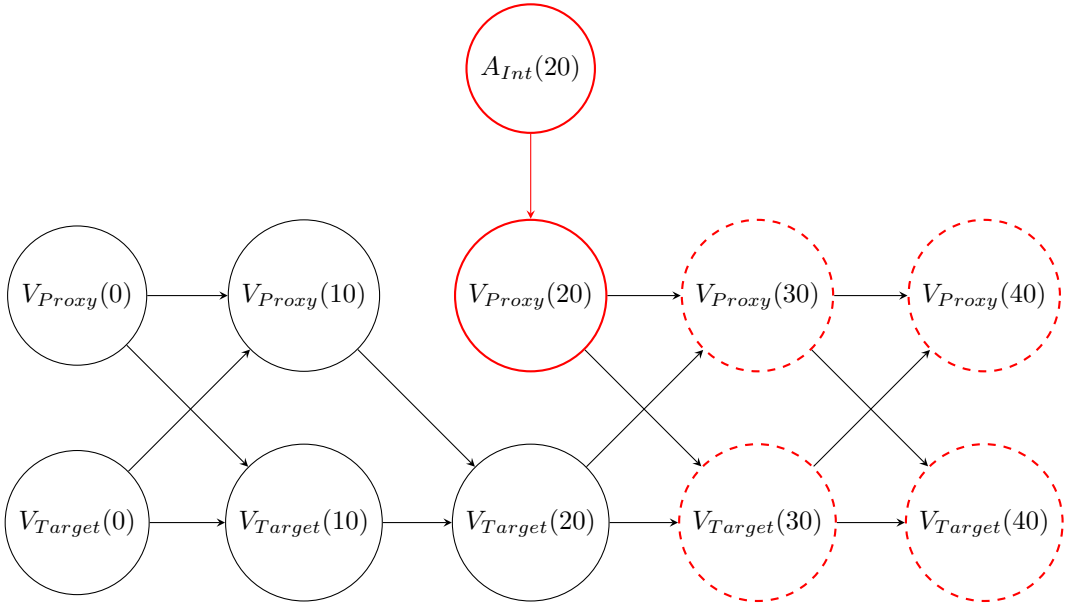
\begin{figure}
    \centering
    
\begin{tikzpicture}[
    >=stealth,
    node distance=3cm,
    dagnode/.style={circle, draw, minimum size=1cm, font=\small}
]

\node[dagnode] (V_P0) {$V_{Proxy}(0)$};
\node[dagnode, right of=V_P0] (V_P10) {$V_{Proxy}(10)$};

\node[dagnode, draw=red!100, thick, right of=V_P10] (V_P20) {$V_{Proxy}(20)$};
\node[dagnode, draw=red!100, thick, above of=V_P20] (A_P20) {$A_{Int}(20)$};

\node[dagnode, draw=red!100, thick=10,dashed, right of=V_P20] (V_P30) {$V_{Proxy}(30)$};
\node[dagnode, draw=red!100, thick=10,dashed, right of=V_P30] (V_P40) {$V_{Proxy}(40)$};

\node[dagnode, below of=V_P0] (V_T0) {$V_{Target}(0)$};
\node[dagnode, below of=V_P10] (V_T10) {$V_{Target}(10)$};
\node[dagnode, below of=V_P20] (V_T20) {$V_{Target}(20)$};
\node[dagnode, below of=V_P30,draw=red!100, thick=10,dashed] (V_T30) {$V_{Target}(30)$};
\node[dagnode, below of=V_P40,draw=red!100, thick=10,dashed] (V_T40) {$V_{Target}(40)$};

\draw[->] (V_P0) -- (V_P10);
\draw[->] (V_P20) -- (V_P30);
\draw[->] (V_P30) -- (V_P40);

\draw[->] (V_T0) -- (V_T10);
\draw[->] (V_T10) -- (V_T20);
\draw[->] (V_T20) -- (V_T30);
\draw[->] (V_T30) -- (V_T40);

\draw[->] (V_T0) -- (V_P10);
\draw[->] (V_T20) -- (V_P30);
\draw[->] (V_T30) -- (V_P40);

\draw[->] (V_P0) -- (V_T10);
\draw[->] (V_P10) -- (V_T20);
\draw[->] (V_P20) -- (V_T30);
\draw[->] (V_P30) -- (V_T40);

\draw[->,draw=red!100,] (A_P20) -- (V_P20);



\end{tikzpicture}

    \caption{    \label{fig:pt_dag_int} Causal Directed Acyclic Graph for $V_{target}(t)$ at different timepoints based on a single intervention on $V_{proxy}(t)$ at $t=20$. The solid red node at $t=20$ is intervened upon, the dashed nodes indicate the nodes after $t=20$ that are changed due to the indirect effects of the intervention. The intervention on $V_{proxy}$ will not immediately affect $V_{target}$ at $t=20$ so the edge is removed.}

\end{figure}

Now consider an intervention designed to cause a change in the volume of the target  cells ($V_{target}$) indirectly via a change to the volume of the proxy cells ($V_{proxy})$.  The intervention can be described using the causal DAG in Figure~\ref{fig:pt_dag_int}.  This intervention would generally be considered as an intervention on a time-varying exposure variable ($V_{proxy}(t)$) to affect the target cell volumes at a particular time-point where past proxy and target cell volumes$(t \geq 30)$ are potential confounders for the present effect of $V_{proxy}(t)$ on $V_{target}(t)$.

What is important to consider in this example is the differentiation between the dynamic causal relationship defined by the feedback parameters and the causal effect of an intervention that would either modify the volume of one of the cells or change the way in which the dynamics of the feedback loop evolve. In the next section, we will explain how to differentiate between these two types of interventions.

\section{Defining Causal Interventions on System Dynamic}
In this section, we propose a framework to rectify the limitations of the standard
causal approaches for defining interventions and the causal effects of interventions. 
We first propose to add an additional criteria to distinguish dynamic null causal effects from the previously described distributional null effects. A dynamic null causal effect exists if there exists a combination of states where the same values of those states will yield the same distribution of outcomes at more than one time point. 
We then propose that  (1) interventions on states of random 
variables that are endogenous to the process (e.g. nodes on the causal DAG and not caused by the outcome) lead to dynamic null causal effects  and (2) interventions on states of random 
variables that are exogenous to the process (i.e. parameters, which can be represented as variables \textit{not} in the causal DAG and not caused by the outcome) lead to dynamic non-null causal effects. We discuss each of these in the next two sections. 

\subsection{Redefinition of the null effect \label{sec:Def}}
 
The classical potential outcome model defines the null causal effect as the average of $Y^{(a)}$
for an intervention $a$ relative to a baseline (or reference) potential 
outcome $Y^{(0)}$, i.e. $E(Y^{(a)}-Y^{(0)})$.  As we saw in the previous section, 
when these potential outcomes are indexed by time ($E(Y_t^{(a)}-Y_t^{(0)})$), the 
choice of $t$ for defining the causal effect becomes critically important. 
Now, readers may find it obvious that causal effects depend on time, but 
the nature of the effect in a dynamic causal model is more nefarious. Consider 
the example in Figure \ref{fig:iso_int} of intervening on a single cell type in an isolated system from the 
previous section. Although the intervention temporarily changes the state
of the process, the dynamics of the system (i.e.the causal relationships between system states) are not altered in a meaningful way and are time-independent. Although the observed trajectories of the system will change after the intervention on a variable's state, the only possible trajectories post-intervention will be trajectories which were possible under the original system dynamics. In short, one observes a change in the system states post-intervention, but the original system dynamics remain unchanged. 

We thus propose instead to differentiate between (1) interventions which lead to 
non-null average causal effects on measurements at specific timepoints and (2) interventions which modify the dynamics and the behaviour of the system over time. The single state example in Figure \ref{fig:iso_int} suggests one way to approach the problem. In this example, an intervention at time $t^*$ affects only the value of $V_{target}(t)$ at specific values of $t>t^*$ and not the underlying 
dynamics of the system.   The classical definition for a causal effect of an intervention $a$ to affect $V_{target}(t)$ is simply $E(V_{target}(t)^{(a)})$ - $E(V_{target(0)}^{(0)})$ $\neq 0$ for \textbf{at least one $t > t_0$}. With time, $E(V_{target}(t))$ may be equal to the same constant equilibrium state as under the null intervention process $V_{target}(t)^{(0)}$ or a different constant equilibrium state value. For any final observation value $T_{max}$, the classical definition of causal effects with potential outcomes, $E(V_{target(T_{max})}^{(a)}) -E(V_{target(T_{max})}^{(0)})$  $\neq 0$ is defined as a causal effect of the intervention on the desired outcome at the final observation time point. 

Figure~\ref{fig:iso_int} shows the difficulty with the classical definition of the causal effect in a dynamic causal system.  The intervention does not actually change the underlying dynamics of the data-generating process; only the values of the state that are observed.  There are only three types of intervention effects one can observe in this system when a state value is changed:  a) achieving the same equilibrium value at an earlier time point; b) achieving the same equilibrium value at a later time point; and c) achieving a different equilibrium value that would be possible under the original time-invariant system dynamics for either a different set of initial states and/or initial time point. For types a) and b), the effects of the interventions are transient.  For type c), the intervention did not change the system equations; only the orbit where the system achieved equilibrium. One way to understand the nature of (c) is that changing only the value of a single state variable in a dynamic system creates a new set of trajectories for all states where current values for unchanged states at the time of the intervention become initial values for the new trajectories with a new initial value for the intervened upon state.  Note that a type c) intervention would be a lasting change to the separate orbits, but would still lead to an orbit achievable without changing the underlying differential equations which govern the process. For example, versions of the Lotka-Volterra model with infinite carrying capacity can have distinct trajectories depending on the initial values (see, for example, Chapter 3 in \citet{murray2002mb1}).  In all three of these cases, the intervention has only affected the states of the process, not the underlying dynamics / causal equations and how the data generating process evolves.

Please note that this paper does not take the position that the three possible causal effects in the previous paragraph are not interesting or potentially valuable; in some situations, these may be the correct effects to consider. However, the classical potential outcome causal effect definition under system dynamics does not allow one to differentiate between interventions on state values and interventions on dynamics. 

Consider instead a different definition of a null potential outcome that allows for more precision in identifying a causal effect on the dynamics of the system. We define the intervention to have a null causal effect on the dynamics of the outcome variable if there exists a combination of a different initial state of the null potential outcome and a different initial time of the null potential outcome process, $\tilde{Y}^{(0)}_{t_0 + \Delta}$, such that 
$E(Y^{(a=1)}_t) -E(\tilde{Y}^{(a=0)}_t) = 0$.  In other words, one can expand the definition of the null potential outcome process to remove identification of effects from interventions which do not affect dynamics of the system.  Figure~\ref{fig:iso_int} shows how an intervention in the previous example can have a non-null effect on $V_{target}(t)$ at particular timepoints, but a null effect on the dynamics. This is because $V_{target}(t)^{(a=1)}$ = $V_{target}(t-10)^{(a=0)}$ for all $t > 10$ under our definition for the one cell volume example. 

Note that our definition also accommodates models with interaction between the two cell types (i.e. $\gamma_{target}$ and $\gamma_{proxy}$ $\neq 0$). For example, an intervention on $V_{proxy}(t)$ at time $t^*$ will induce a perturbation to the $V_{target}$, but will not change how the two systems evolve together.  In other words, the $V_{target}(t)^{(a=1)}$ = $V_{target}(t+\Delta)^{(a=0)}$ for all $t > t^*$ if the intervention creates a situation where the target and proxy cell volumes are identical to volumes that could have been observed with no intervention but at a different initial time $t+\Delta$, i.e. $V_{proxy}(t)^{(a=1)}$ = $V_{proxy}(t+\Delta)^{(a=0)}$ and $V_{target}(t)^{(a=1)}$ = $V_{target}(t+\Delta)^{(a=0)}$.  Therefore, the same three types of effects on the $V_{target}$ are the only effects possible: (1) increased time to equilibrium, (2) decreased time to equilibrium or (3) a different equilibrium under the same dynamics.  

We propose to use a generalization of the null potential outcome to allow for different initial states of both $\tilde{V}_{proxy}(t+\Delta)^{(a=0)}$ and $\tilde{V}_{target}(t+\Delta)^{(a=0)}$, but where $\Delta$ is the same for both processes.

\begin{figure}
    \centering
\begin{subfigure}[l]{\textwidth}
    \includegraphics{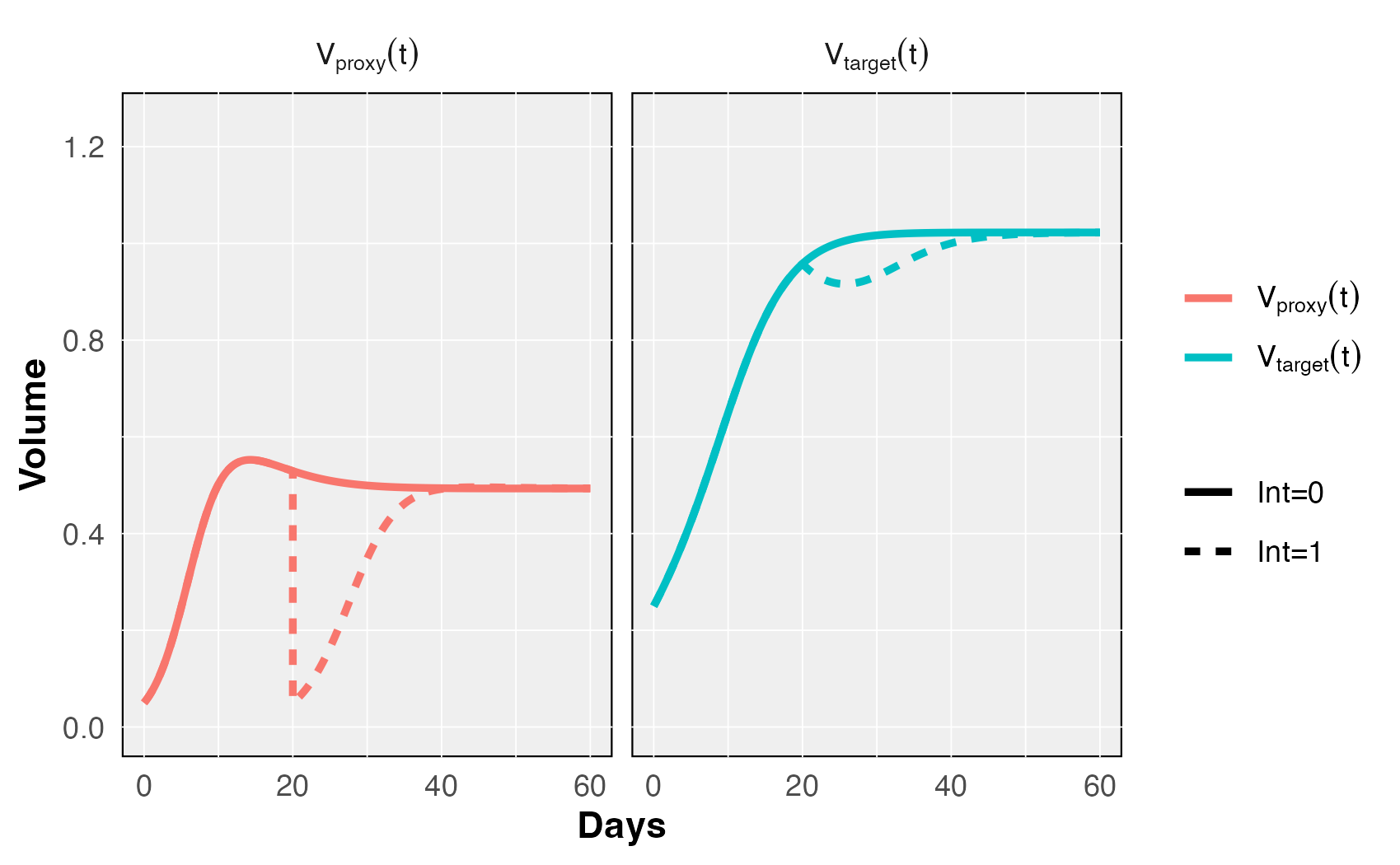}
    \caption{ Figure (a) left panel shows $V_{proxy}(t)$ when there is no intervention (solid line) and under an intervention (dotted line) at time $t=20$ to change the value of $V_{proxy}(t)$ to its initial value. Figure (a) right panel shows $V_{target}(t)$ for no intervention on $V_{proxy}(t)$ (solid line) and the change in $V_{target}(t)$ after intervening on $V_{proxy}(t)$ at time $t=20$ (dotted line).}. 
\end{subfigure}
\\
\begin{subfigure}[l]{\textwidth}
    \centering
    \includegraphics{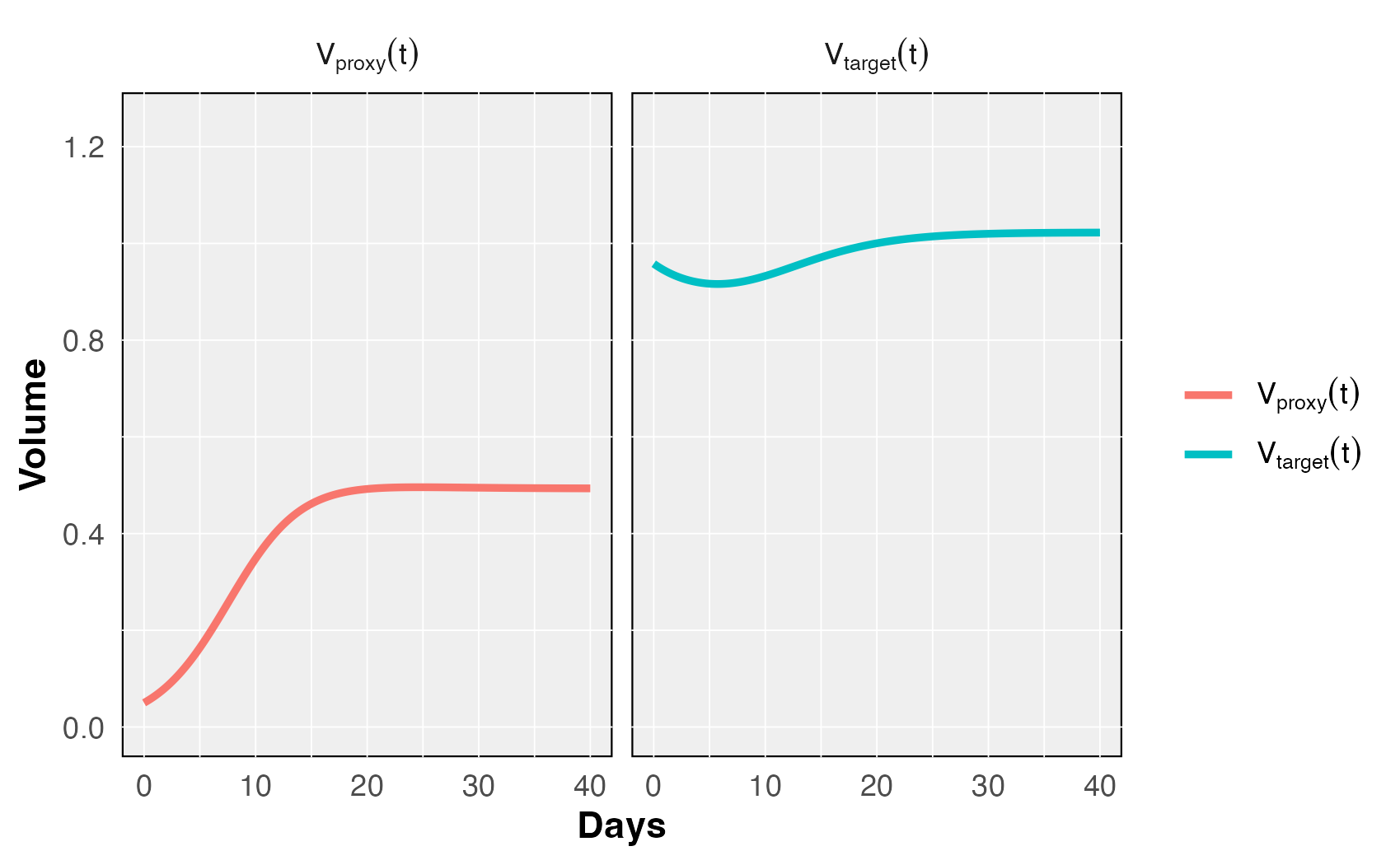}
    \caption{ $V_{proxy}(t)$ and $V_{target}(t)$ trajectories from Day 0 when initial states are equal to $V_{proxy}(20)$ and $V_{target}(20)$ from panel (a) and there is no intervention. Note that the trajectories in panel (a) from $t=20$ are the same those in panel (b) for $\Delta=-20$ from time point $t=20+\Delta=0$. } 
\end{subfigure}

\caption{ \label{fig:comp_noint_null} Examples of intervening on systems with competition.}

\end{figure}

In the next subsection, we explain what types of interventions are necessary to achieve non-null effects on system dynamics under this new definition of the null potential outcome.

\subsection{Intervening on the dynamics of a system \label{sec:intervene}}

In the previous subsection we proposed a definition of a new null potential outcome which would only allow for causal effects that modify system dynamics, as opposed to just affecting the values of state variables. 

The critical issue in our illustrative example (and in most dynamic systems with feedback loops) is the endogeneity of the variable that is being intervened upon. In particular, the values of the parameters, not the states, determine the dynamics and equilibrium behaviour of the system.  The key difference between the parameters of the system and the states of the system in Equation \ref{eq:LVeq_int}  is that none of the parameters depend on the state values, which means the parameters can be considered to be exogenous to the process and the causal DAG.  Because the parameters are exogenous, intervening directly on the parameters could induce changes to the dynamics of the system. In other words, by changing the values of the parameters, you create post-intervention trajectories which could not have been obtained under the original dynamics. 

\begin{figure}

    \centering
\begin{subfigure}[l]{\textwidth}
    \includegraphics{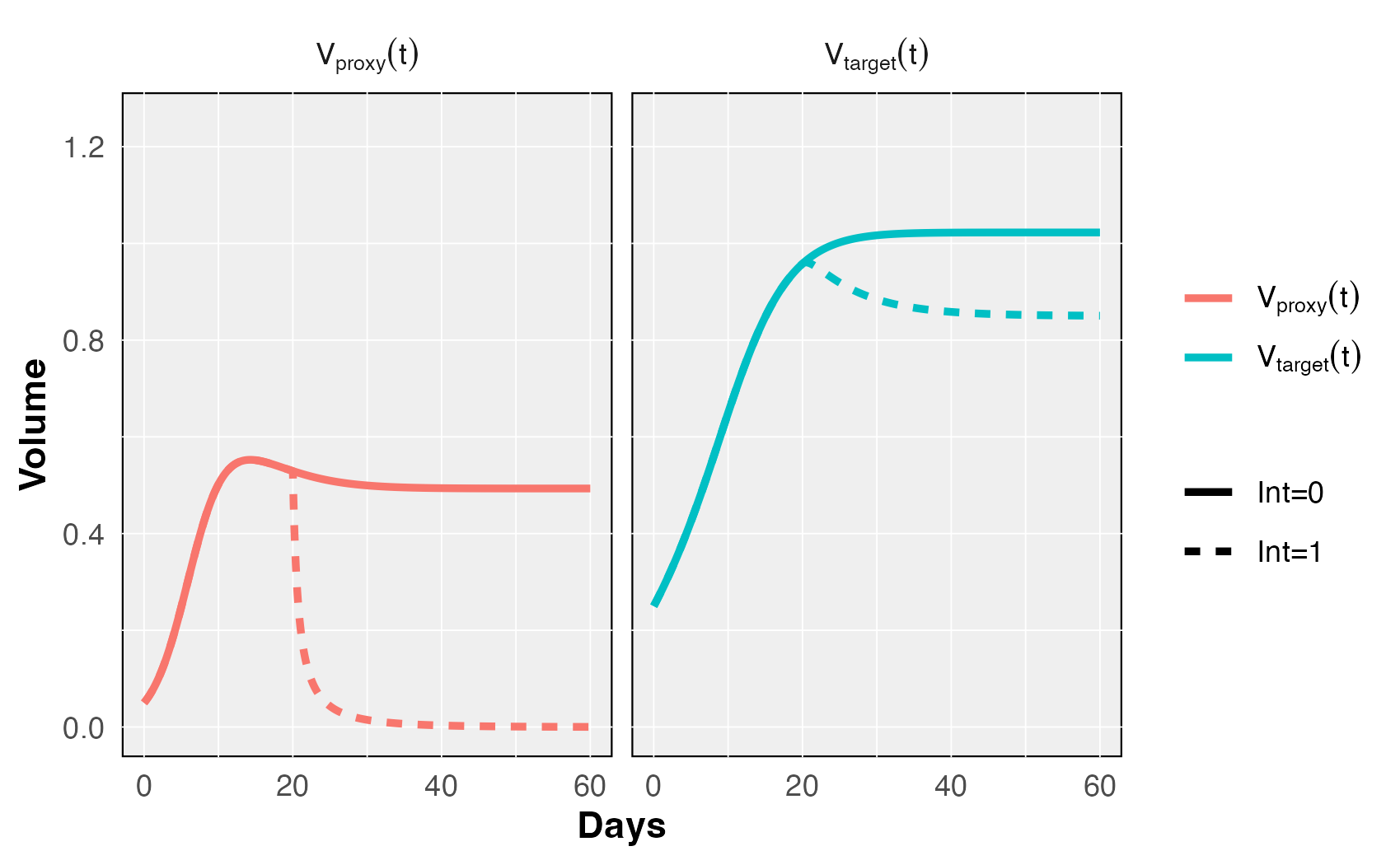}
    \caption{ Figure (a) left panel shows $V_{proxy}(t)$ when there is no intervention (solid line) and under an intervention at time $t=20$ to change the value of $K_{proxy}$ from 0.8 to 0.2. Figure (a) right panel shows $V_{target}(t)$ for no intervention on $V_{proxy}(t)$ (solid line) and the change in $V_{target}(t)$ after intervening on $K_{proxy}$ at time $t=20$.\label{fig:comp_noint_null_a}}
        \label{fig:par_sub_a}
\end{subfigure}
\\
\begin{subfigure}[l]{\textwidth}
    \centering
    \includegraphics{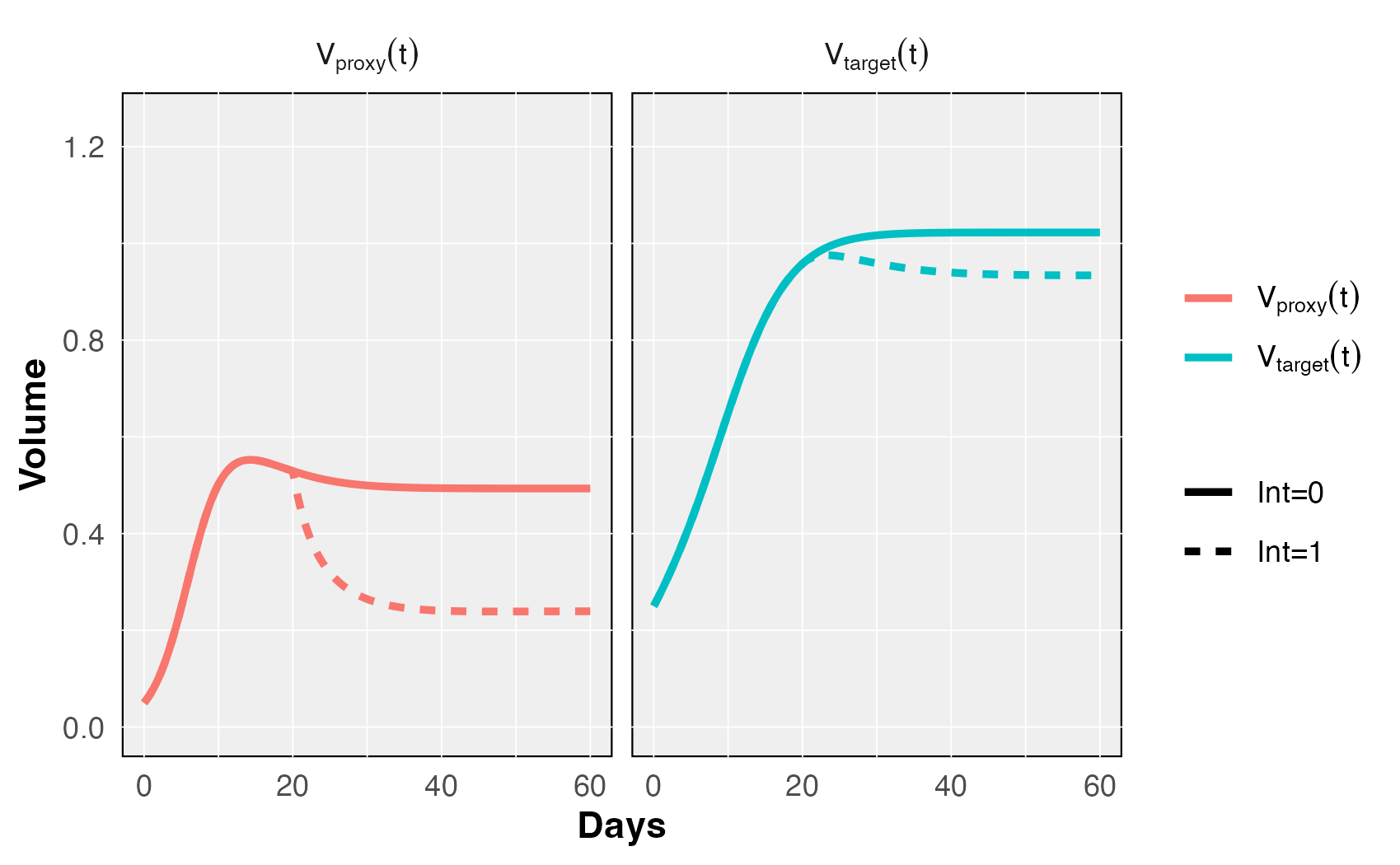}
    \caption{ Figure (b) left panel shows $V_{proxy}(t)$ when there is no intervention (solid line) and under an intervention at time $t=20$ to change the value of $\gamma_{proxy}$ from 0.3 to 0.6. Figure (b) right panel shows $V_{target}(t)$ for no intervention on $V_{proxy}(t)$ (solid line) and the change in $V_{target}(t)$ after intervening on $K_{proxy}$ at time $t=20$.\label{fig:comp_noint_null_b}}
        \label{fig:par_sub_b}
\end{subfigure}

\caption{Example of two interventions that lead to different system dynamics \label{fig:comp_int_par}}

\end{figure}

Figure \ref{fig:comp_int_par} illustrates the effects of two possible interventions on the data generating process for $V_{proxy}(t)$: (a) one intervention on the carrying capacity ($K_{proxy}$) of the $Type_{proxy}$ cells and (b) one intervention on the interaction parameter ($\gamma_{proxy}$) which governs how $V_{proxy}(t)$ affects the rate of change of $V_{target}(t)$. Note that in both cases we obtain processes after the intervention that are different from before the intervention (i.e. $t=0$) even when the values of the variables are identical to $t=0$.

For the purposes of illustration, we show that intervening on parameters of the system rather than endogenous states allow us to obtain causal effects on system dynamics. Clearly defining an intervention on a model parameter deviates from the mainstream causal practice.\footnote{Targeting interventions to parameters is a key component of our approach. For interventions to have an effect on anything, they must cause a change in the physical universe in which we live. This is straightforward when discussing measured variables on the causal DAG. What does it mean to change a parameter, and how can we do so? In brief, a causal DAG does not usually include all causes of all variables. Rather, it only has to include common causes of the exposure of interest and outcome. Consider that A causes Y faster if B is present (e.g. an enzyme), or only if C is present. In Rothman's sufficient causal set framework,\citep{rothman1995causes} both \{A,B\} and \{A,C\} represent sufficient causal sets for the arrow from A to Y, even though B and C (referred to as component causes) may not be on the causal DAG. If we intervene on B (or C), we will change the magnitude of the causal effect of A on Y, i.e. we are intervening on a variable in our physical universe that changes the parameter.}
However, if a state variable is exogenous in that it causes changes in the variables endogenous to the dynamic system, but is not itself caused by variables endogenous to the dynamic system, then it plays the same role as a parameter with respect to the data generating mechanism of the dynamic system.

\section{Relationship to other research}

In this section, we describe how our research connects and contrasts to relevant work done in other areas of dynamic causal modelling. 

\subsection{Directed graphs for equilibrium models}

\begin{figure}
    \centering

\begin{tikzpicture}[
   >=stealth,
    node distance=3cm,
    dagnode/.style={circle, draw, minimum size=1cm, font=\small}
]



\node[dagnode,color=blue] (parm) {$\gamma_{Target}$};
\node[dagnode,color=blue,below of=parm] (parm2) {$K_{Proxy}$};

\node[dagnode, right of=parm] (dV_P) {$dV_{Proxy}$};
\node[dagnode,color=red,right of=dV_P] (V_P) {$V_{Proxy}$};

\node[dagnode, below of=dV_P] (dV_T) {$dV_{Target}$};
\node[dagnode, below of=V_P] (V_T) {$V_{Target}$};

\draw[->] (V_T) -- (dV_P);

\draw[->] (V_P) -- (dV_T);

\draw[<-,dashed] (V_P) -- (dV_P);
\draw[<-,dashed] (V_T) -- (dV_T);


\draw[->]  (V_T) to[bend left] (dV_T);
\draw[->]  (V_P) to[bend left=-30] (dV_P);
\draw[->]  (parm) to (dV_P);
\draw[->]  (parm2) to (dV_P);

\end{tikzpicture}

 \caption{\label{fig:naft_dag} Dynamic causal graph for equations (\ref{eq:LVeq_int_A}) and (\ref{eq:LVeq_int}) using the notation of \citet{iwasaki1994causality, weinberger2023intervening}.  Intervening on the state value of the red node leads to a change in state with no change to the dynamics of the system as described in Section~\ref{sec:Def}.  Intervening on the parameters (blue nodes) leads to a change in the dynamics of the system as described in Section~\ref{sec:intervene}.}
    
\end{figure}
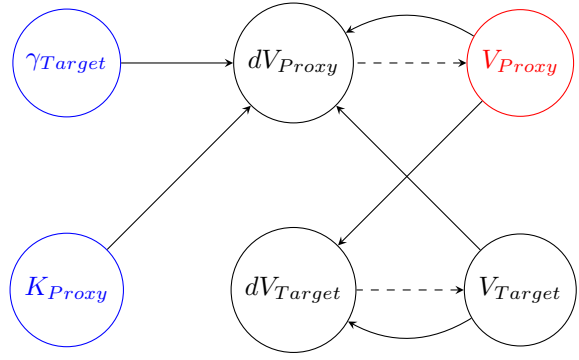

\citet{iwasaki1994causality} made one of the earliest contributions for characterizing causal effects (in particular, causal ordering) for equilibrium models via directed effects. Their work defines causal ordering in self-contained dynamic systems and generalizes the idea of casusal ordering to systems which contain multiple self-contained dynamic systems.   They also discuss the role of exogeneity in terms of how the casual graph should treat variables that are either unaffected or weakly unaffected by other variables in the system, at least with regards to the particular time scale.  The causal graphs proposed in their paper include derivatives as variables in the directed graph. In Figure~\ref{fig:naft_dag}, we give an example of how their work applies to the dynamic model for the cells in Equations \ref{eq:LVeq_int_A} and \ref{eq:LVeq_int}.

\citet{weinberger2023intervening} builds on the work of Iwasaki and Simon to further characterize causal effects for equilibrium models. In particular, he resolves a key issue when trying to learn 
causal graphs from data obtained from models at equilibrium. He highlights that a dynamical system can end up in distinct equilibrium states with distinct causal orderings depending on whether one either (1) intervenes to ``clamp'' a variable at a particular value, or (2) allows the variable to reach equilibrium as a result of its endogenous dynamics. Each of the equilibrium models resulting from each policy can be derived from the corresponding dynamic model, which includes derivatives for variables that have not reached equilibrium.  Our paper directly extends this work to the potential outcome framework and clarifies the ways in which one can intervene on derivatives in a dynamic causal graph. Interventions on derivatives are distinct from clamp interventions, since they do not fix the values of state variables, but they also bring the system to a different equilibrium state than would result from its endogenous dynamics. 

\citet{blom2023causality} have also contributed work related to the use of directed graphs for inferring the absence (or presence) of perfectly adapted systems, i.e. systems that return to perfect adaptation after having been perturbed away from that equilibrium behaviour.  In their approach, they use the conditional independence structure estimated from the graph to determine whether the intervention prevented perfect adaptation or not.  Their approach would be the hypothesis testing version of our attempt to estimate effects on the basis of our novel definition of potential outcomes.

\subsection{Connections to work for stochastic differential models}

In our paper we have focused on the use of ordinary differential equations, but recent work 
by \citet{boeken2024dynamic} defines a version of do-calculus and directed (both cyclic and acyclic) graphs for {\em stochastic} differential equations. Like our work in this paper, the authors
emphasize the importance of differentiating between endogenous and 
exogenous state variables.  They do not define potential outcomes, and causal relationships
between state variables are inferred via the absence and presence of edges (similar to \citet{blom2023causality}). 

\subsection{Related approaches for functional causal effects}

There has been other recent work done in the causal literature using similar definitions of a functional potential outcome. \citet{ecker2024causal} define a Functional Average Treatment Effect (the FATE) in a similar way as we have in Section \ref{sec:charact}. In their example, they illustrate a comparison between two treatment groups (early adult residence location, urban or rural) with respect to a functional outcome (log cumulative income). They then define 
an individual potential outcome as $\theta(t) = E[y_{1i}(t)-y_{0i}(t)]$ for the function that defines the functional causal effect of an urban early adult residence location relative to a rural location.  Their treatment of interest is exogenous in the way it is defined, so their context would not require the novel potential outcome definition of causal effects on the dynamics.  Our work would allow one to consider the possibility of a changing residence location which is also functional.  

\citet{testa2025doubly} define the same functional treatment effect as a target for 
their proposed new doubly robust methods for functional estimation. Similar to \citet{ecker2024causal}, they only consider a binary, exogenous intervention, so there is no need for the more complex potential outcome that we propose in this paper. 

\subsection{Connections to dynamic treatment regimes}

Finally, dynamic treatment regime models have been proposed for problems with 
time-varying exposures where the goal is to estimate the causal effect of a 
sequence of interventions \citep{murphy2003optimal,robins2004optimal}. Our potential outcome approach for dynamic models would
mirror that of the dynamic treatment potential outcomes in that we could allow
for a sequence of interventions at different timepoints 
$(a_{T_0}, a_{T_1}, ... , a_{T_k})$. The same two issues for the single intervention 
case would apply for the multiple intervention case, but would be more complex.  
First, we would need 
to define the reference null potential outcome relative to what could have occurred 
at potentially different initial values under the same dynamics after each 
intervention.  Second, the standard dynamic treatment regime models do not differentiate
between endogenous and exogenous variables, so the same issue with respect to 
affecting local states but not system dynamics would still apply.

\subsection{Connections to differences-in-differences approaches}

Differences-in-differences (DiD) methods provide the closest intuition to the potential outcome models we present in our paper.  The basic DiD model applied to our tumour cell volume example would compare the pre-post difference in the rate of change in the target tumour cell growth with the intervention to the pre-post difference in the rate of change in target tumour cell growth without the intervention, while adjusting for potential confounding factors.  \citep{angrist2009mostly,card1993minimum}. The DiD null causal effect definition depends on an assumption that the trajectories of the two potential outcomes are parallel post-intervention (e.g. allowing for different intercepts, but the same slope in a linear model). This is similar to our proposed null effect for dynamic systems which assumes the same trajectory for systems with and without interventions with the one trajectory being just a shifted version of the other in time. 


\section{Discussion and future work}

Our paper makes two key contributions for identifying relevant causal effects on the dynamics of systems defined by ordinary differential equations.  The first contribution characterizes a more general approach to specifying functional null potential outcomes in order to be able to differentiate between interventions that causally influence state values and interventions that causally influence system dynamics.   The second contribution of our paper is to provide guidance to researchers on how to understand what parts of the system can and cannot be intervened upon in order to produce causal effects on system dynamics, namely that interventions designed to affect system dynamics should target mechanisms that are exogenous to the dynamical system.

Our paper focuses on dynamical systems defined by homogeneous ordinary differential equations as such models allow for clear illustration of the difficulties in identifying causal effects on the dynamics of systems. Generalizing to non-homogeneous ordinary or partial differential equation models would make the characterization of the set of null potential outcomes more difficult, but would not be intractable. Moving to other types of systems with less hypothesized structure would also be possible, but increasing the flexibility of the functional form would likely require more data observed with higher temporal resolution.

We have not discussed estimation in the current paper, as we wanted to focus only on the causal identification problem. Causal estimation using the generalized form of the null potential outcome is not necessarily straightforward, even under randomized interventions. In traditional causal estimation, estimates of null (or reference) potential outcomes are generally estimated from observations before (or without any) intervention. For randomized interventions with no confounding, one approach would be to attempt to perform functional registration \citep{ramsay2005functional,ramsay2007parameter} to compare trajectories for intervened upon sample units to reference unit trajectories at potentially different timepoints with different starting values.  If a parametric ODE can be used to model the data reasonably well, G-computation approaches could be used estimate intervention trajectories under different initial values which would allow closer comparison of system dynamics with reference trajectories.  Such methods would work well for trajectories with appropriate frequency of observed measurements; such an estimation strategy would likely work less well with irregularly spaced or infrequent observations over time.  When the interventions are not randomly assigned to observations, one would need to then account for confounding of the intervention in the usual way, either through adjustment or weighting (or both).  We hope to pursue all these directions in our future work on this problem.

\bibliography{DCP}

\end{document}